\documentclass[aps,prl,twocolumn,secnumarabic]{revtex4-1}
    \usepackage{graphicx}
    \usepackage{amsmath}
    \usepackage{amsfonts}
\usepackage{setspace}

\begin{document}
        
        \title{Finite-system Multicriticality at the Superradiant Quantum Phase Transition}
        \author{Han-Jie Zhu$^{1}$}
        \author{Kai Xu$^{1}$}
        \author{Guo-Feng Zhang$^{1}$}
        \email{gf1978zhang@buaa.edu.cn}
        \author{Wu-Ming Liu$^{2,3,4}$}
        \affiliation{$^{1}$Key Laboratory of Micro-Nano Measurement-Manipulation and Physics (Ministry of Education), School of Physics, Beihang University, Xueyuan Road No. 37, Beijing 100191, China}
        \affiliation{$^{2}$Beijing National Laboratory for Condensed Matter Physics, Institute of Physics, Chinese Academy of Sciences, Beijing 100190, China}
        \affiliation{$^{3}$School of Physical Sciences, University of Chinese Academy of Sciences, Beijing 100190, China}
        \affiliation{$^{4}$Songshan Lake Materials Laboratory, Dongguan, Guangdong 523808, China}

        \begin{abstract}
We demonstrate the existence of finite-system multicriticality in a qubit-boson model where biased qubits collectively coupled to a single-mode bosonic field. 
The interplay between biases and boson-qubit coupling produces a rich phase diagram which shows multiple superradiant phases and phase boundaries of different orders. 
In particular, multiple phases become indistinguishable in appropriate bias configurations, which is the signature of multicriticality. 
A series of universality classes characterizing these multicritical points are identified. 
Moreover, we present a trapped-ion realization with the potential to explore multicritical phenomena experimentally using a small number of ions. 
The results open a novel way to probe multicritical universality classes in experiments.
        \end{abstract}
        \maketitle

Quantum multicriticality, where multiple phases simultaneously become identical at a specific quantum critical point, 
is a fascinating phenomenon as well as fundamental concept in the study of quantum phase transitions \cite{Quantum_Phase_Transitions}. 
At a multicritical point, the system is governed by a new universality class, 
which result in qualitatively different critical behaviors including new scaling fields and critical exponents \cite{Scaling_and_Renormalization_in_Statistical_Physics}. 
Owing to this unique nature, 
intriguing features and novel universality classes have been found in various multicritical systems, 
including magnetic materials \cite{Quantum_tricriticality_in_antiferromagnetic_Ising_model_with_transverse_field}, 
superconductors \cite{Proximity_of_iron_pnictide_superconductors_to_a_quantum_tricritical_point}, 
optical systems \cite{Emergent_Universality_in_a_Quantum_Tricritical_Dicke_Model} and 
various condensed matter systems \cite{Chiral_Tricritical_Point,Tricriticalities_and_Quantum_Phases_in_Spin-Orbit-Coupled_Spin_1_Bose_Gases,Quantum_Tricriticality_at_the_Superfluid_Insulator_Transition_of_Binary_Bose_Mixtures,
Phase_diagram_and_multicritical_behaviors_of_mixtures_of_three_dimensional_bosonic_gases,Quantum_Multicriticality_near_the_Dirac_Semimetal_to_Band_Insulator_Critical_Point_in_Two_Dimensions,
Fermionic_multicriticality_near_Kekule_valence_bond_ordering,Itinerant_quantum_multicriticality_of_two_dimensional_Dirac_fermions,Quantum_multicriticality_in_disordered_Weyl_semimetals}.

Despite the novelty and significance, 
the investigation of quantum multicriticality is still very limited due to enormous challenges in experiments. 
To reach a high-order critical point, multiple parameters need to be finetuned precisely, 
e.g., accurate adjustments of composition and magnetic field are both required to access the tricritical point in $\mathrm{Nb}_{1-y} \mathrm{Fe}_{2+y}$ \cite{Quantum_tricritical_points_in_NbFe2}. 
This imposes a much stricter requirement on the controllability of actual systems compared to normal critical cases \cite{Quantum_tricritical_points_in_NbFe2,Reentrant_Superconductivity_Driven_by_Quantum_Tricritical_Fluctuations_in_URhGe,Magnetic_field_induced_antiferromagnetic_tricritical_points_in}. 
Furthermore, while universality class plays a central role in multicritical phenomena, its exploration is even more difficult since universal behavior emerges only when the system size is sufficiently large, 
and such behavior is vulnerable to environmental noises owing to the long preparation time of groundstate caused by the critical slowing down.
To reveal a universality class, 
we need to maintain the controllability of a large-size system while preserving universal behavior from noise effects, 
which is extreme difficult in realistic settings \cite{Protecting_quantum_information_against_environmental_noise}.

Instead of entering the large-size limit, 
a finite-size system may also undergo a quantum phase transition (QPT) if the thermodynamic limit can be reached in an alternative way. 
This is the case of the Dicke model \cite{Coherence_in_Spontaneous_Radiation_Processes,Nonequilibrium_Quantum_Phase_Transitions_in_the_Dicke_Model,Dissipation_Induced_Anomalous_Multicritical_Phenomena,
Introduction_to_the_Dicke_Model,Single_Photon_Triggered_Quantum_Phase_Transition,Universal_Dissipationless_Dynamics,Unified_superradiant_phase_transitions,Critical_phenomena_in_an_extended_Dicke_model,Scrambling_in_the_Dicke_model,Spectrum_of_the_Dicke_model_in_a_superconducting_qubit_oscillator_system,Superradiant_phase_transition_in_the_ultrastrong_coupling_regime,
Universal_Spectral_Features_of_Ultrastrongly_Coupled_Systems,Critical_Quantum_metrology_with_a_finite_component_quantum_phase_transition}, 
which describes a bosonic mode collectively coupled to multiple qubits.
In this model, a second-order QPT appears
when the ratio of the mode frequency to the qubit transition frequency approaches zero \cite{Quantum_Phase_Transition_and_Universal_Dynamics_in_the_Rabi_Model,Quantum_Phase_Transition_in_the_Finite_Jaynes_Cummings_Lattice_Systems,Universal_Scaling_and_Critical_Exponents_of_the_Anisotropic_Quantum_Rabi_Model}. 
This model can be realized in different systems ranging from ultracold atoms \cite{Dicke_quantum_phase_transition_with_a_superfluid_gas_in_an_optical_cavity,Realization_of_the_Dicke_Model_Using_Cavity_Assisted_Raman_Transitions,
Verification_of_a_Many_Ion_Simulator_of_the_Dicke_Model_Through_Slow_Quenches_across_a_Phase_Transition,
Measuring_out_of_time_order_correlations_and_multiple_quantum_spectra_in_a_trapped_ion_quantum_magnet,Cold_atom_based_implementation_of_the_quantum_Rabi_model,Nonlinear_quantum_Rabi_model_in_trapped_ions} 
to superconducting circuits \cite{Experimentally_simulating_the_dynamics_of_quantum_light_and_matter_at_deep_strong_coupling,
A_tunable_Josephson_platform_to_explore_many_body_quantum_optics_in_circuit-QED,Microwave_photonics_with_superconducting_quantum_circuits,Superradiant_Phase_Transition_in_a_Superconducting_Circuit},
where some of them, e.g., trapped-ion systems, have shown the possibility of achieving finite-system QPT due to its excellent controllability in the required critical regime  \cite{Probing_the_Dynamics_of_a_Superradiant_Quantum_Phase_Transition_with_a_Single_Trapped_Ion,Quantum_Simulation_of_the_Quantum_Rabi_Model_in_a_Trapped_Ion,Onset_of_a_quantum_phase_transition_with_a_trapped_ion_quantum_simulator,
Analog_quantum_simulation_of_generalized_Dicke_models_in_trapped_ions,Exploring_nonequilibrium_phases_of_the_generalized_Dicke_model}. 
For such systems, an interesting question arise: can multicriticality be induced while maintaining the system size small? 
If so, this would be highly desirable since it enables the study of multicriticality in a system with sufficient controllability and noise suppression ability due to small system size. 
More interestingly, is it possible to explore universality classes 
through probing the critical behavior under realistic conditions?

In this Letter, we show the existence of quantum multicritical points in a finite-size qubit-boson model via engineering qubit biases. 
We show that qubit biases can introduce novel features in the phase diagram, 
and multicritical points emerge in certain bias configuration. 
These points can be characterized by a series of multicritical universality classes. 
Critical exponents and scaling relations describing these universality classes are also obtain. 
Finally, we consider a trapped-ion realization with the potential to explore multicritical phenomena experimentally using a small number of ions. 
The numerical results show that it is possible to correctly reveal universality classes at multicritical points 
by non-equilibrium universal functions even with noise effects.

\begin{figure}
            \centering
            \includegraphics[scale=1]{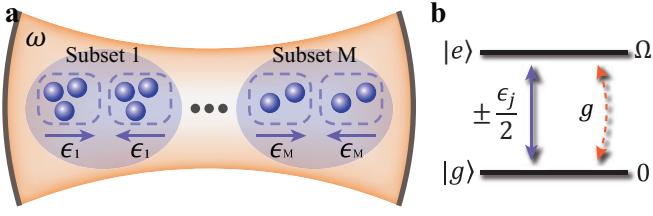}
            \caption{
               (a) A schematic illustration of the qubit-boson model with staggered bias configuration (Eq. \ref{Hamiltonian}). 
Here a bosonic field with frequency $\omega$ is collectively coupled to $2N$ qubits (blue spheres) which are seperated into $M$ subsets. 
Within each subset, qubits are divided into two halves and each half experiences a transverse field with the same magnitude and opposite direction to the other.
(b) The level scheme of a qubit within the $j$th subset. 
The qubit couples to the bosonic field with interaction strength $g$, and an additional bias $\pm \epsilon_j /2$ is presented due to the transverse field. 
            }
            \label{Fig0}
        \end{figure}

\emph{Model and phase diagram.}---We consider a bosonic field coupled to qubits with staggered bias configuration (Fig. \ref{Fig0}), characterized by the Hamiltonian ($\hbar=1$)
\begin{eqnarray}
\label{Hamiltonian}
H=&&\,\omega a^{+}\! a\!+\!\sum_{j=1}^{M}\left[\frac{\Omega}{2}\left(J_{z, 2 j-1}\!+\!J_{z, 2 j}\right)\!+\!\frac{\epsilon_{j}}{2}\left(J_{x, 2 j-1}\!-\!J_{x, 2 j}\right)\right] \nonumber\\
&&+\frac{g}{\sqrt{2 N}} \sum_{j=1}^{M}\left(J_{x, 2 j-1}+J_{x, 2 j}\right)\left(a^{+}+a\right),
\end{eqnarray}
where  $a^+$ ($a$) is the creation (annihilation) operator of the bosonic field with frequency $\omega$. 
Here $2N$ qubits are split into $M$ subsets,  
and for each subset qubits are further divided into two halves which are equipped with biases with the same magnitude 
($\epsilon_{j}$ for the $j$th subset with qubit number $2N_j$) 
but opposite signs. 
The collective spin operators $\boldsymbol{J}_{2 j-1}=\sum_{i=1}^{N_{j}} \boldsymbol{\sigma}_{j}^{(i)} / 2$ and 
$\boldsymbol{J}_{2 j}=\sum_{i=1}^{N_{j}} \boldsymbol{\sigma}_{j}^{\left(N_{j}+i\right)} / 2$ 
are composed of the Pauli operators $\boldsymbol{\sigma}_{j}^{(i)}$ describing the $i$th qubit within the $j$th subset, 
$\Omega$ and $g$ are the energy spacing of qubits and qubit-boson interaction strength respectively. 
In the absent of biases, the system returns to the original Dicke model and undergoes a second-order QPT 
in the $\omega / \Omega \rightarrow 0$ or $N \rightarrow \infty$ limit \cite{Quantum_Phase_Transition_and_Universal_Dynamics_in_the_Rabi_Model}. 
The biases can be introduced in various realizations of the Dicke model, 
e.g., by applying effective transverse Zeeman fields to atoms or ions systems \cite{Boson_mediated_quantum_spin_simulators_in_transverse_fields}, 
or tuning the persistent currents in superconducting qubits \cite{Microwave_photonics_with_superconducting_quantum_circuits}. 

\begin{figure}
            \centering
            \includegraphics[scale=1]{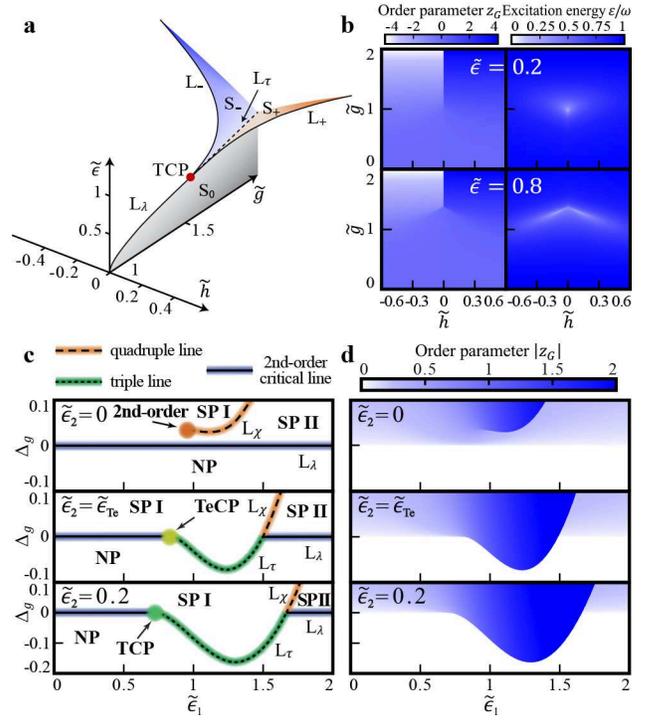}
            \caption{
               (a)  Phase diagram in the tricritical case ($M=1$). 
The solid lines and dash line represent the second-order critical lines and triple line respectively. 
The first-order coexistence surfaces $S_0$ separates two superradiant phases with different order parameter sign, 
while $S_\pm$ separate a normal phase and two superradiant phases respectively.
(b) The groundstate order parameter $z_G$ (left panels) 
and the excitation energy $\varepsilon$ (right panels) as functions of $\tilde{g}$ and $\tilde{h}$.
(c) Phase diagram in the tetracritical case ($M=2$ and $N_1=3N_2$), 
where $\Delta_{g}=\tilde{g}-\tilde{g}_{r}(\tilde{\epsilon}_1,\tilde{\epsilon}_2)$ is the distance to $L_\lambda$ for given $\tilde{\epsilon}_{1,2}$.  
$\tilde{g}_{r}(\tilde{\epsilon}_1,\tilde{\epsilon}_2)$ is the coupling value on $L_\lambda$ when $\tilde{\epsilon}_{1,2}$ are fixed, and is the solution to $r(\tilde{g} _r, \tilde{\epsilon}_1,\tilde{\epsilon}_2)=0$ 
since $L_\lambda$ is determined by $r=0$. 
(d) The groundstate order parameter $|z_G|$ for fixed $\tilde{\epsilon}_2$.
            }
            \label{Fig1}
        \end{figure}

The biases can introduce important novel features in the phase diagram. 
To investigate the phase structure, we first resort to the MF approach 
and the ground-state properties can be analyzed by minimizing the energy functional per qubit (see the Supplemental Material \cite{Supplimentary_material} for details)
\begin{equation}
E(z)\!=\!\frac{z^{2}}{4 \tilde{g}^{2}}\!-\!\frac{1}{4} \sum_{j=1}^{M} n_{j}(\sqrt{\left(z\!+\!\tilde{\epsilon}_{j}\right)^{2}\!+\!1}\!+\!\sqrt{\left(z\!-\!\tilde{\epsilon}_{j}\right)^{2}\!+\!1})
\end{equation}
where $n_{j}=N_{j} / N$ is the number fraction of the $j$th subset, 
$\tilde{g}=2 g / \sqrt{\omega \Omega}$ and $\tilde{\epsilon}_{j}=\epsilon_{j} / \Omega$ 
are the dimensionless coupling strength and bias respectively, 
and $z=2 \sqrt{\eta} \tilde{g} \varphi$ is the rescaled order parameter characterizing the superradiant transition 
where $\varphi=\langle a\rangle$ and $\eta=(2 N \Omega)^{-1} \omega$ is the frequency ratio. 
The superradiant QPT is marked by a transition from normal phase (NP) ($\varphi=0$) to superradiant phase (SP) ($\varphi \neq 0$), 
and the corresponding critical points form a manifold in the parameter space. 
In this manifold, multicritical points of at most ($M+2$)-th order can arise, 
which can be shown by expanding $E(z)$ up to ($2M+4$)-th order of $z$ as 
$E(z)=E_{0}+v\!\left(r z^{2} / 2\!+\!\sum_{j=1}^{M} u_{j} z^{2(j+1)} /(2 j\!+\!2)\!+\!z^{2(M+2)} /(2 M\!+\!4)\right)$. 
For appropriate $\left\{n_{j}\right\}$ settings, 
it is possible that the coefficients $r$ and $u_{1}, \dots, u_{M}$ vanish simultaneously 
since there exists $M+1$ independent parameters $\tilde{g}$ and $\left\{\tilde{\epsilon}_{j}\right\}$. 
This point is nothing but an $(M+2)$-th order critical point if it further satisfies $v>0$. 
For a complete description of this multicritical point, we further introduce 
symmetry-breaking biases $H_{n s}\!=\!\sum_{j} h_{j}\left(J_{x, 2 j-1}\!+\!J_{x, 2 j}\right) / 2$ to the Hamiltonian. 
The resulting energy functional $E_{ns}(z)$ can be expanded as 
$E_{n s}\left(z\right)\!=\!E\left(z_{n s}\right)\!+\! v \sum_{j=1}^{M+1} w_{j} z_{n s}^{2 j-1} /(2 j-1)$ 
up to $O({\tilde{h}_{j}})$, 
where $\tilde{h}_{j}=h_{j} / \Omega$ and $z_{n s}=z-z_{0}$ 
with the constant $z_0$ chosen to remove the $z_{n s}^{2 M+3}$ term \cite{Supplimentary_material}. 
Then at this critical point, $r$, $\left\{u_{j}\right\}$ and $\left\{w_{j}\right\}$ form a complete set of scaling variables, 
and the critical behavior can be described in terms of these variables.

The simplest $M=1$ case permits the appearance of tricritical points (TCP).
Fig. \ref{Fig1}(a) presents the extended phase diagram in the parameter space $(\tilde{g}, \tilde{\epsilon}, \tilde{h})$ (subscripts of $\tilde{\epsilon}_1$ and $\tilde{h}_1$ are omitted in this case). 
Here the second-order critical line $L_\lambda$ turns into a triple line $L_\tau$ 
where three phases coexist as the bias $\tilde{\epsilon}$ is strong enough, 
and their meeting point additionally connects to two wing critical lines $L_\pm$. 
The critical lines $L_\lambda$ and $L_\pm$ further connect to the coexistence surfaces $S_0$ and $S_\pm$ respectively. 
These structures are the signatures of tricriticality \cite{Theory_of_tricritical_points}, and the meeting point is a TCP whose location can be determined by 
$u=r=0$, i.e., $\left(\tilde{g}_{T}, \tilde{\epsilon}_{T}\right)=\left((5 / 4)^{3 / 4}, 1 / 2\right)$. 
The occurrence of tricriticality can also be observed through the groundstate order parameter, as shown in Fig. \ref{Fig1}(b). 
Clearly, the tricriticality causes a bifurcation of the first-order surfaces $S_0$ into two wings $S_\pm$ ended at $L_\pm$. 

The next case is $M=2$, which allows the existence of tetracritical points (TeCP) where four phases become identical simultaneously. 
For $N_1=3N_2$ case, the system features tetracriticality as shown in Fig. \ref{Fig1}(c, d), 
which illustrates the phase diagram and the corresponding order parameter $z_G$ for fixed $\tilde{\epsilon}_{2}$. 
When $\tilde{\epsilon}_{2}$ is small, a new pair of SPs with opposite order parameters emerges. 
These new phases indicate that the interplay between biases and boson-qubit coupling 
induces different collective behaviors of qubits. 
Different SP pairs are separated by a first-order quadruple line $L_{\chi}$ where four SPs coexist, 
and at its endpoint the QPT turns into second-order where the difference between two SP pairs vanishes. 
As $\tilde{\epsilon}_{2}$ is increased to a specific value $\tilde{\epsilon}_{2, T e}$, 
this end point will finally reach the second-order critical line $L_\lambda$ determined by $r=0$. 
After this value, the endpoint stays on $L_\lambda$ and turns into a TCP which connects to a triple line $L_\tau$. 
Thus these endpoints form a quadruple line and a tricritical line when 
$\tilde{\epsilon}_{2}<\tilde{\epsilon}_{2, T e}$ and $\tilde{\epsilon}_{2}>\tilde{\epsilon}_{2, T e}$ respectively, 
and at the meeting point all four SPs become indistinguishable, 
which signifies the appearance of tetracriticality. 
Its location can be determined numerically as $\left(\tilde{g}_{T e}, \tilde{\epsilon}_{1, T e}, \tilde{\epsilon}_{2, T e}\right) \approx(1.30,0.81,0.15)$ 
through the equation $u_{1}=u_{2}=r=0$.
For $M>2$ cases, higher order critical points are possible.
For example, a pentacritical point, where the NP and four SPs simultaneously become identical, 
exists at $\left(\tilde{g}_{P}, \tilde{\epsilon}_{1, P}, \tilde{\epsilon}_{2, P}, \tilde{\epsilon}_{3, P}\right) \approx(1.36,0.98,0.37,0.17)$ 
when $N_{1}=4 N_{2}=4 N_{3}$.

\begin{figure}
            \centering
            \includegraphics[scale=1]{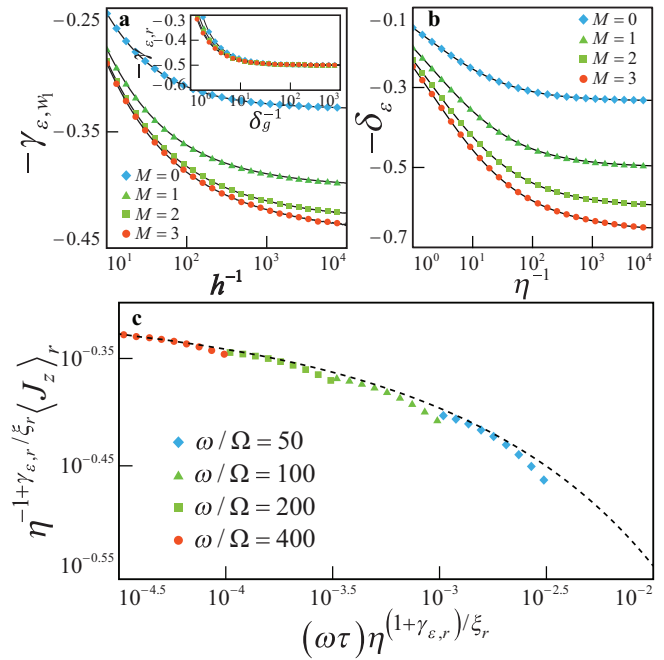}
            \caption{
			(a) Fits of the critical exponents $\gamma_{\varepsilon, w_{1}}$ and $\gamma_{\varepsilon, r}$ (inset of (a)) at critical points of different orders. 
(b) Fits of the critical exponent $\delta_{\varepsilon}$.
(c) Numerical results of the residual qubit population $\eta^{-1+\gamma_{\varepsilon, r} / \xi_{r}}\left\langle J_{z}\right\rangle_{r}$ 
after quench as a function of $\tau \eta^{\left(1+\gamma_{\varepsilon, r}\right) / \xi_{r}}$ for different frequency ratio with noise effects included. 
The black dash line is the non-equilibrium scaling function $\mathcal{S}_{J_{z}}$ obtained in the $\eta \rightarrow 0$ limit numerically.
            }
            \label{Fig2}
        \end{figure}
\emph{Critical behavior and universality.}---Next we consider the critical behavior of the ground-state order parameter $z_G$. 
At an ($M+2$)-th order critical point ($\tilde{g}_{(M)},\left\{\tilde{\epsilon}_{j,(M)}\right\}$), 
the system should belong to a different universality class compared to the lower-order case and a new scaling relation emerges with different critical exponents. 
The critical behavior of $z_G$ can be obtained by the equation $\partial E_{n s} / \partial z=0$, which leads to the scaling relation as 
$z_{G}=w_{1}^{\beta_{w_{1}}} \mathcal{M}_{n s, z}\left(\left\{x_{\mathcal{A}}\right\}_{ \mathcal{A}=r,\left\{u_{j}\right\},\left\{w_{j}\right\}_{j \neq 1}}\right)$ 
where $x_{\mathcal{A}}=\mathcal{A} w_{1}^{-\phi_{w_{1}, \mathcal{A}}}$ and 
$\mathcal{M}_{n s, z}$ is the scaling function (the subscript $ns$ denotes the non-symmetry case $\left\{h_{j}\right\} \neq 0$) \cite{Supplimentary_material}. 
The exponent $\beta_{\mathcal{A}}$ describes the singularity related to the variable $\mathcal{A}$ near the critical point and can be calculated as 
$\beta_{r}=1 /(2 M+2)$, $\beta_{u_{j}}=1 /(2 M-2 j+2)$ and $\beta_{w_{j}}=1 /(2 M-2 j+5)$, 
while $\phi_{\mathcal{A}_1,\mathcal{A}_2}=\beta_{\mathcal{A}_1}/ \beta_{\mathcal{A}_2}$ is the crossover exponent \cite{crossover_exponent}. 
This relation suggests that the leading singularity of $z_G$ is path-dependent as moving toward the multicritical point, 
and is given by $z_{G} \sim{\{\tilde{h}_{j}^{\beta_{w_{1}}}\}}$ 
except approaching from the direction with $w_1=0$. 
This is due to $\phi_{w_{1},\mathcal{A} }<1$ for $\mathcal{A} \neq w_{1}$ 
and the scaling variables behave as $r \sim \delta_{\left(g,\left\{\epsilon_{j}\right\}\right)}$, $u_{j} \sim \delta_{\left(\left\{\epsilon_{j}\right\}\right)}$ 
and ${w_{j} \sim \{\tilde{h}_{j}\}}$ near the multicritical point, 
where $\delta_{g}=\tilde{g}-\tilde{g}_{(M)}$ and $\delta_{\epsilon_{j}}=\tilde{\epsilon}_{j}-\tilde{\epsilon}_{j,(M)}$. 
In the symmetry case ($\left\{h_{j}\right\} = 0$), the scaling relation reduces to 
$z_{G}=|r|^{\beta_{r}} \mathcal{M}_{z}\left(\left\{u_{j}|r|^{-\phi_{r, u_{j}}}\right\}\right)$ when $r\le0$ (superradiant phases), 
while $z_G$ vanishes in the $r>0$ region (normal phase). 
The leading singularity is $z_{G} \sim \delta_{\left(g,\left\{\epsilon_{j}\right\}\right)}^{\beta_{r}}$ 
for all directions with $r \neq 0$ since $\phi_{r, u_{j}}<1$ for all $u_j$.

We now include quantum fluctuations and investigate further critical behavior. 
In the $\omega / \Omega \rightarrow 0$ limit, the low-lying energy states behaves as a harmonic oscillator, 
and the effective Hamiltonian can be written as 
$H_{e f f}=\varepsilon a^{+} a+C$ where $\varepsilon$ is the excitation energy and $C$ is a constant \cite{Supplimentary_material}. 
This excitation energy vanishes at second- and higher-order critical points 
since $\varepsilon$ satisfies $\varepsilon^{2} \propto \partial^{2} E_{n s} /\left.\partial z^{2}\right|_{z=z_{G}}$ 
and the r.h.s. is zero at these critical points (the second- or higher-order nature of these critical points ensures the vanishing of quadratic terms in the expansions of $E_{n s}$ at $z=z_{G}$). 
This can also be demonstrated in Fig. \ref{Fig1}(c), which shows the closing of the energy gap near the critical lines $L_\lambda$ and $L_\pm$. 
Furthermore, near a ($M+2$)-th order critical point, the relation between $\varepsilon$ and $\partial^{2} E_{n s} / \partial z^{2}$ provides the scaling relation 
$\varepsilon=\left|w_{1}\right|^{\gamma_{\varepsilon, w_{1}}} \mathcal{M}_{n s, \varepsilon}\left(\left\{x_{\mathcal{A}}\right\}\right)$,  
where $\gamma_{\varepsilon, w_{1}}=(M+1) /(2 M+3)$ \cite{Supplimentary_material}. 
Thus the leading singularity is $\varepsilon \sim \{\tilde{h}_{j}^{\gamma_{\varepsilon, w_{1}}}\}$ 
for directions with $r \neq 0$. 
In Fig. \ref{Fig2}(a), we show fits of $\gamma_{\varepsilon, w_{1}}$ which are obtained numerically, and it converges to the analytical values as approaching the critical point.
Similarly, in the symmetry case, 
the scaling relation reduces to 
$\varepsilon=|r|^{\gamma_{ \varepsilon, r}} \mathcal{M}_{\varepsilon}\left(\left\{u_{j}|r|^{-\phi_{r, u_{j}}}\right\}\right)$. 
The leading singularity is $\varepsilon \sim r^{\gamma_{ \varepsilon, r}} \sim \delta_{\left(g,\left\{\epsilon_{j}\right\}\right)}^{\gamma_{\varepsilon, r}}$ 
for all directions with $r \neq 0$, 
and an $M$-independent exponent $\gamma_{\varepsilon, r}=1 / 2$ presents (inset of Fig. \ref{Fig2}(a)). 

The difference between these critical points can be further revealed by considering the finite-frequency scaling, 
which describes the emergence of critical behavior as $\eta$ approaches zero. 
The results show that at the critical point, the excitation energy vanishes as $\varepsilon \sim \eta^{\delta_{\varepsilon}}$ 
where $\delta_{\varepsilon}=\gamma_{\varepsilon, r} / \xi_{r}=\gamma_{\varepsilon, w_{1}} / \xi_{w_{1}}$ is the finite-size scaling exponent, 
and $\xi_{r}=(M+3) /(2 M+2)$, $\xi_{w_{1}}=(M+3) /(2 M+3)$ are
observable-independent exponents which are specific to the universality class (Fig. \ref{Fig2}(b)) \cite{Supplimentary_material}. 
Apparently, multicritical points with different orders are indeed belong to different universality classes 
with distinct scaling fields and critical exponents. 
These classes are the extensions of the Dicke universality class $M=0$ in the multicritical regime.

\emph{Experimental realization.}---In principal, 
we can confirm the multicritical nature of a critical point via estimating critical exponents experimentally. 
However, this approach is practically very difficult since it requires an adiabatic preparation of groundstate near a critical point, 
which requires significantly long time due to the vanishing energy gap. 
The coherence time is in general much shorter than this preparation time, 
thus the critical behavior will be severely distorted by environmental noise. 
Instead, the non-equilibrium scaling function is much robust under environmental noises due to lower time requirement 
\cite{Probing_the_Dynamics_of_a_Superradiant_Quantum_Phase_Transition_with_a_Single_Trapped_Ion,New_Dynamical_Scaling_Universality_for_Quantum_Networks,
Symmetry_Breaking_Bias_and_the_Dynamics_of_a_Quantum_Phase_Transition}. 
Therefore, we consider a linear quench $\tilde{g}(t)=\tilde{g}_{(M)} t / \tau$ 
in the symmetry case while fixing the biases $\tilde{\epsilon}_{j}=\tilde{\epsilon}_{j,(M)}$, 
where $\tau$ is the duration of quench. 
The system is initially prepared in the groundstate, and we focus on the residual qubit population 
$\left\langle J_{z}\right\rangle_{r} \equiv\left|\left\langle J_{z}\right\rangle_{f}(\eta, \tau)-\left\langle J_{z}\right\rangle(\eta)\right|$ 
at the end of the quench since $\left\langle J_{z}\right\rangle$ can be measured with high fidelity in the trapped-ion setup \cite{Quantum_Simulation_of_the_Quantum_Rabi_Model_in_a_Trapped_Ion}. 
Here $\left\langle J_{z}\right\rangle_{f}(\eta, \tau)$ and $\left\langle J_{z}\right\rangle(\eta)$ 
denote the expectation value of $J_z$ after quench and that of the groundstate when $t=\tau$ respectively. 
When the quench is sufficiently slow, the majority of excitations are produced inside the critical regime 
and $\left\langle J_{z}\right\rangle_{r}$ satisfies a scaling relation \cite{Supplimentary_material}
\begin{equation}
\left\langle J_{z}\right\rangle_{r}=\eta^{1-\gamma_{\varepsilon, r} / \xi_{r}} \mathcal{S}_{J_{z}}\left(\tau \eta^{\left(1+\gamma_{\varepsilon, r}\right) / \xi_{r}}\right),
\end{equation}
where $\mathcal{S}_{J_{z}}$ is the non-equilibrium scaling function. 
If $\eta^{-1+\gamma_{\varepsilon, r} / \xi_{r}}\left\langle J_{z}\right\rangle_{r}$ 
is plotted as a function of $\tau \eta^{\left(1+\gamma_{\varepsilon, r}\right) / \xi_{r}}$, 
all data points with different $\eta$ should collapse to a single curve, 
which allows us to determine the order of a critical point and reveal its universality class via $\xi_{r}$.

This approach is possible in experiments. 
For simplicity, we focus on the tricritical case, and consider an experimental realization comprised of two trapped ions which are cooled down to their motional groundstates. 
Here qubits are encoded using different electronic states \cite{Quantum_simulation_of_the_Dirac_equation}, 
while the bosonic field are the center-of-mass vibrational mode supported by the Coulomb repulsion and confining potentials 
\cite{Effective_Quantum_Spin_Systems_with_Trapped_Ions}. 
The qubit biases can be generated by additional near-resonant lasers \cite{Boson_mediated_quantum_spin_simulators_in_transverse_fields}. 
Finally, the spin-phonon coupling is induced by a pair of laser beams with frequencies slightly detuned from the red- and blue-sideband respectively 
\cite{Quantum_Rabi_Model_with_Trapped_Ions,Boson_mediated_quantum_spin_simulators_in_transverse_fields,Quantum_Simulation_of_the_Quantum_Rabi_Model_in_a_Trapped_Ion}. 
In this setup, the system can be described by an effective Hamiltonian which has the desired form Eq. (\ref{Hamiltonian}) with $N=M=1$, 
and the parameters associate with the experimental ones as $\omega=\left(\delta_{b}-\delta_{r}\right) / 2$, 
$\Omega=\left(\delta_{b}+\delta_{r}\right) / 2$, $g=\sqrt{2} \eta_{0} \Omega_{0}$ and $\epsilon=\Omega_{p}$ \cite{Supplimentary_material}. 
Here $\delta_{b}$ ($\delta_{r}$) is the detuning to the blue- (red-) sidebands, 
$\Omega_{0}$ and $\eta_{0}$ are the Rabi strength and Lamb-Dicke parameter of the blue/red-sideband lasers respectively, 
and $\Omega_{p}$ is the Rabi strength of the laser which produces staggered biases. 
For typical trapped-ion platforms, it is possible to achieve $\omega=(2 \pi) \, 200\, \mathrm{H z}$ and 
frequency ratios $50 \leq \Omega / \omega \leq 400$ \cite{Quantum_simulation_of_the_Dirac_equation}. 
To reach the tricritical point, it is necessary to realize the Rabi frequencies $9.9 \leq \Omega_{0} / (2 \pi) \leq 27.9\, \mathrm{k H z}$ and
$5.0 \leq \Omega_{p} / (2 \pi) \leq 40.0 \, \mathrm{kHz}$ ($\eta_{0}=0.06$ is considered), 
which are attainable in experiment \cite{Quantum_simulation_of_the_Dirac_equation}.

We now evaluate whether the scaling function can be correctly retrieved when noise effects are taken into account. 
Here we only consider phonon heating as the main noise source since the qubit dephasing produced by the magnetic-field fluctuations 
can be effectively suppressed via continuous dynamical decoupling \cite{A_robust_scheme_for_the_implementation_of_the_quantum_Rabi_model_in_trapped_ions,
Protected_ultrastrong_coupling_regime_of_the_two_photon_quantum_Rabi_model_with_trapped_ions,Magnetic_field_fluctuations_analysis_for_the_ion_trap_implementation_of_the_quantum_Rabi_model}, 
and the qubit decay is much slower than the phonon heating rate 
($\sim 1\, \mathrm{s}$ qubit lifetime vs. $\sim 100\, \mathrm{m s}$ phonon coherence time) \cite{The_coherence_of_qubits_based_on_single_Ca_ions}. 
Then the quench process is governed by a phenomenological master equation 
$\dot{\rho}=-i[H(\tilde{g}(t)), \rho]+\gamma\left(n_{\mathrm{th}}+1\right) \mathcal{D}[a]+\gamma n_{\mathrm{th}} \mathcal{D}\left[a^{+}\right]$ 
where $\mathcal{D}[A]=A \rho A^{+}-\left\{\rho, A^{+} A\right\} / 2$, 
$\gamma$ is the phonon-reservoir coupling strength and $n_{\mathrm{th}}$ is the mean phonon number when the system is in equilibrium with the reservoir.
The phonon heating rate is set to a typical value $\gamma n_{\mathrm{th}} / \omega=0.05$ 
and the effective temperature of the reservoir is high enough such that $\gamma n_{\mathrm{th}} \approx \gamma\left(n_{\mathrm{th}}+1\right)$. 
Fig. \ref{Fig2}(c) shows the quench results of $\left\langle J_{z}\right\rangle_{r}$ with the noise effects, 
where the quench time is chosen in a range $0.75 \leq \omega \tau \leq 2$ ($3.75 \leq \tau \leq 10\, \mathrm{m s}$) 
which is much shorter than the phonon coherence time. 
We can find that all data points with different $\eta$ collapse into a theoretical line of $\mathcal{S}_{J_{z}}$ approximately, 
thus the scaling function $\mathcal{S}_{J_{z}}$ can be faithfully retrieved under environmental noises. 
This allows identification of multicritical universality classes.

\emph{Conclusion.}---In conclusion, we have shown that finite-system multicriticality can be induced by the interplay between qubit biases and boson-qubit coupling.
In certain bias configurations, 
multiple phases become indistinguishable and this relates to a high-order critical point 
which resides between a multiple coexistence line and a lower-order critical line in the phase diagram. 
These points can be characterized by a series of multicritical universality classes. 
Moreover, we have presented a trapped-ion realization with the potential to explore multicritical phenomena experimentally. 
Due to the small system size, we are able to retain necessary controllability and coherence under realistic conditions, 
thus make it possible for experiments to reveal the multicritical universality classes through non-equilibrium universal functions. 
Our work extends the multicriticality study to finite-size systems, and provide a promising platform for experimental exploration.

        \bibliography{FinitesystemMulticriticality.bbl}

\begin{thebibliography}{63}
\bibitem [{cro()}]{crossover_exponent}%
  \BibitemOpen
  \href@noop {} {}\bibinfo {note} {The scaling function $\mathcal{M}_{n s, z}$
  satisfies $\lim\limits_{x_{\mathcal{A}} \rightarrow \infty} \mathcal{M}_{n s,
  z} \sim x_{\mathcal{A}}^{\beta_{\mathcal{A}}}$ and remaines finite when all
  $x_{\mathcal{A}}$ approach zero. Therefore,
  $\phi_{\mathcal{A}_1,\mathcal{A}_2}$ characterizes the crossover between
  different scaling behaviors related to the variables $\mathcal{A}_1$ and
  $\mathcal{A}_2$. For example, in the $M=0$ case, the scaling relation of
  $z_{G}$ is given by $z_{G}=w_{1}^{\beta_{w_{1}}} \mathcal{M}_{n s, z} (r
  w_{1}^{-\phi_{w_{1}, r}})$. Then we have $z_{G} \sim w_{1}^{\beta_{w_{1}}}$
  when $|r| \ll |w_{1}|^{\phi_{w_{1}, r}}$, while $z_{G} \sim r^{\beta_r}$ when
  $|r| \gg |w_{1}|^{\phi_{w_{1}, r}}$. Since $\phi_{w_{1}, r}<1$, the leading
  singularity is $z_{G} \sim w_{1}^{\beta_{w_{1}}}$ when the critical point is
  approached from nearly all direction in the parameter space. The only
  exception is from the special direction with $w_{1}=0$. In this case the
  leading singularity is $z_{G} \sim r^{\beta_r}$. Similar argument can also be
  applied to the $M>0$ case and other scaling functions.}\BibitemShut {Stop}%
\end{thebibliography}

    \end{document}